\documentclass[epj]{svjour}
\usepackage{graphicx}
\begin{document}
\title{Quantitative Tube Model for Semiflexible Polymer Solutions}
\author{Hauke Hinsch \inst{1} \and Jan Wilhelm \inst{1} \and Erwin
  Frey \inst{1}
}                     
\offprints{frey@lmu.de}          
\institute{ Arnold Sommerfeld Center for Theoretical Physics and
  Center of NanoScience,\\
  Department of Physics, Ludwig-Maximilians-Universit\"at M\"unchen, \\
  Theresienstrasse 37, D-80333 M\"unchen, Germany}
\date{Received: date / Revised version: date}
%
\abstract{We develop a analytical and quantitative theory of the tube
  model concept for entangled networks of semiflexible polymers. The
  absolute value of the tube diameter $L_\perp$ is derived as a
  function of the polymers' persistence length $l_{\rm p}$ and mesh
  size $\xi$ of the network. To leading order we find $L_\perp = 0.32
  \xi^{6/5} l_{\rm p}^{-1/5}$, which is consistent with known
  asymptotic scaling laws.  Additionally, our theory provides
  corrections to scaling that account for finite polymer length
  effects and are dominated by the mesh size to polymer length ratio.
  We support our analytical studies by extensive computer simulations.
  These allow to verify assumptions essential to our theoretical
  description and provide an excellent agreement with the analytically
  calculated tube diameter. Furthermore, we present simulation data
  for the distribution function of tube widths in the network.}

\maketitle
\section{Introduction}
Filamentous actin (F-actin) is a semiflexible biopolymer that has been
the object of intensive research from several domains. As a major
constituent of the cytoskeleton, F-actin networks play a key role in
the ongoing puzzle of cell mechanics \cite{howard01,bausch06} and cell
motility \cite{pantaloni01}. Depending on the presence of binding
proteins, F-actin strands at medium concentration can form both
chemical (cross-linked) and physical (entangled) networks with
different elastic properties \cite{gardel04,wagner06,hinner98}.
Besides rheological methods, single polymer visualizations also are
feasible, as the strands exceed most synthetic polymers by length.
This facilitated the observation of tube-like regions along which DNA
or F-actin filaments reptate \cite{perkins94,kas94}. The confinement
of polymers to these cylindrical cages confirmed the ``tube model''
postulated earlier by de Gennes \cite{gennes79} and Doi and Edwards
\cite{doi86}.  This long standing paradigm had proven a successful
concept to reduce the complex structure of entangled networks to a
single polymer problem.

\begin{figure}[htbp]
\center
\includegraphics[width=\columnwidth]{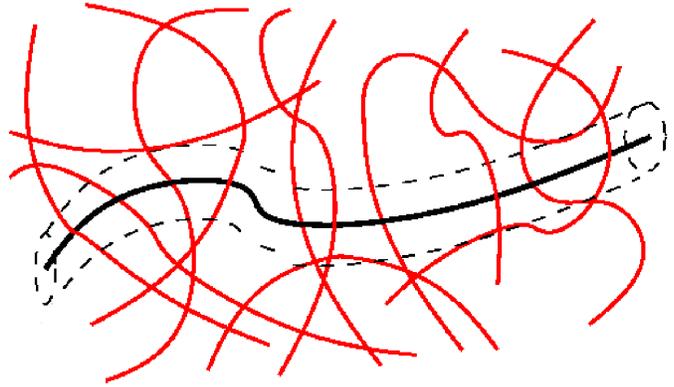}
\caption{The effect of all surrounding polymers that hinder the test
  polymer's transverse displacement is described by a hypothetical
  tube.}
\label{fig:cartoon}
\end{figure}

In such entangled networks, polymers can effortlessly slide past each
other but are not allowed to cross. Their interaction is thus mainly
of entropic nature as entanglements mutually restrict the accessible
configuration space. Grasping this feature in a single polymer model
has lead to the famous tube model \cite{gennes79,doi86}. The
suppression of transverse undulations of a test polymer by the
surrounding polymers (Fig.\ref{fig:cartoon}) is modelled by a tube.
This tube follows the average path of the test polymer and its profile
is frequently modelled by a harmonic potential. The average strength
of this potential is determined by the local density of the network.
The tube concept has proven a successful tool to derive scaling laws
for several network properties \cite{doi86,odijk83}. For example, due
to the confinement energy of the filament inside, the tube diameter
can be connected to mechanical properties of the network, e.g. the
different moduli \cite{hinner98,mackintosh95,isambert96}. However, due
to the phenomenological nature of the tube model, most of its benefits
have been mainly qualitative.  Recently, also quantitative predictions
of the plateau modulus and the tube diameter of flexible polymers
melts were achieved by a novel approach based on the microscopic
foundations and the topological structure of the network
\cite{everaers04,tzoumanekas06}.

Even if most concepts developed for flexible polymers can not be
carried over to the semiflexible case with its large persistence
length, the tube model is perfectly applicable as well. However, while
in general scaling laws of the tube diameter \cite{semenov86} or the
plateau modulus \cite{isambert96} are well established, quantitative
theories are still under debate and lack from approval by measurements
of sufficient accuracy. Again the challenge is to make the successful
tube model quantitative by connecting the phenomenological tube and
its microscopic origins. In the present work, we contribute to the
discussion by supplying an absolute value for the tube diameter from a
theory supported by extensive computer simulations.

We will proceed as follows: in Section \ref{sec:model-definition} the
model under investigation is defined and all relevant length scales
are discussed. By the analysis of the free energy cost for confining a
polymer to a hypothetical tube, the tube diameter is derived as a
function of Odijk's deflection length for finite-length polymers. The
appropriate deflection length for a given polymer concentration and
persistence length is derived in the following sections. To this end,
the polymer is modeled by a sequence of independent rods in Section
\ref{sec:indep-rod-model}. Criteria for the correct choice of the
independent rod length and a self-consistent determination of the tube
diameter are developed, before a final result for the tube diameter is
obtained in Section \ref{sec:plugging-it-all}. Extensive numerical
simulations supporting these result and providing additional insight
are presented in Section \ref{sec:simulations} followed by our
conclusion in Section \ref{sec:conclusion}.

\section{Model Definition}\label{sec:model-definition}
We consider a mono-disperse network of physically entangled polymers
with a particular focus on pure solutions of the biopolymer F-actin.
The polymer density is given by the number $\nu$ of polymers of length
$L$ per unit volume. The polymers are of bending stiffness $\kappa$
corresponding to a persistence length $l_{\rm p}=\kappa/k_{\rm B} T$.
A single polymer's configuration ${\bf r}(s)$ is parameterized by the
arc length $s$ and the average distance between the polymer chains can
be characterized by a mesh size $\xi:=\sqrt{3/(\nu L)}$ \footnote{The
  mesh size has the unit length and can be interpreted as an average
  distance between network constituents. While the denominator
  $\sqrt{1/\nu L}$ ensures the correct scaling the numerator is a mere
  definition.}. We will describe the constituent polymers by the
worm-like chain model \cite{kratky49,saito67} and exploit the tube
model concept \cite{gennes79,doi86} to reduce the description of the
network to a single polymer and its neighbors. In the following we
will begin our analysis with an investigation of the different length
scales involved in the system.

\subsection{Length Scales}\label{sec:length-scales}
Typical F-action solutions are polydisperse with a mean filament
length $L \approx 22 \mu$m \cite{kaufmann92}. With a persistence
length $l_{\rm p} \approx 17 \mu$m \cite{gittes93,goff02} comparable
to its length, it is the textbook example of a semiflexible polymer.
At a concentration of $c=0.5$~mg/ml corresponding to $\nu \approx 1
\mu m^{-3}$ \cite{schmidt89} the average mesh size equals $\xi
\approx 0.4 \mu$m. We can thus state that the persistence length of a
filament is much larger than the distance to its neighbors, $l_{\rm p}
\gg \xi$.  Since the tube diameter $L_\perp$ is at most of the order
of the mesh size, this additionally implies $l_{\rm p} \gg L_\perp$.
The polymer will thus not deviate far from the tube center.
Consequently, configurations where the polymer folds back onto itself
are rendered unlikely. This is a minimal requirement to model the tube
by a harmonic potential of strength $\gamma$. The potential has to be
seen as a hypothetical tube representing the joint contribution of all
surrounding polymers which constrain the transverse undulations of a
given polymer (see Fig.\ref{fig:cartoon}).

The energy of a certain polymer contour ${\bf r}(s)$ is the sum of the
bending energy of the polymer and its confinement into the harmonic
potential and is given in the weakly-bending rod approximation by
\begin{equation} \label{eq:hamilton}
H(\gamma,\kappa)=\int_0^L ds \left[\frac{\kappa}{2}({\bf
    r}_\perp^{\prime \prime}(s))^2+\frac{\gamma}{2}{\bf r}^2_\perp(s)
\right] \;.
\end{equation} 
Here ${\bf r}(s)=(s,{\bf r}_\perp(s))$ is a parameterization in
arc-length $s$ and transverse displacement ${\bf
  r}_\perp(s)=(y(s),z(s))$ from the tube center. The prime denotes a
derivative with respect to $s$. This harmonic approximation to the
Hamiltonian of the worm-like chain model is valid as long as $\vert
{\bf r}_\perp^{\prime \prime} \vert \ll 1$, i.e. as long as the
transverse coordinates of the tube coordinate can be considered to
remain single valued.

With the thermal average $\langle \cdot \rangle$ the tube diameter can now
be defined as
\begin{equation}
L_\perp := \frac{1}{L} \, \big\langle \int_0^L ds \, 
               {\bf r}_\perp^2 (s) \big\rangle \;.
\end{equation}
So far we have identified two length scales: the length scale of
persistence length and the total polymer length describing the
properties of one specific polymer, and the length scale of mesh size
and the tube diameter describing the properties of the network
structure.  Additionally we introduce the deflection length $L_{\rm d}
:= (\kappa/\gamma)^{1/4}$ as a third useful length scale. It is
interpreted below as that length on which interactions between single
polymer and network occur. More precisely, it is a measure for the
number of contacts of the polymer with the tube walls. For large
confinement strength $\gamma$ the tube is small, making interaction
with the encaged polymer more likely and therefore resulting in a
small deflection length. On the other hand, for a large polymer
rigidity $\kappa$ transverse undulations allowing contacts with the
tube walls are energetically unfavorable and the distance between
contact will decrease.  For $l_{\rm p} \gg L_\perp$ we expect the
deflection length to be distinctively smaller than the polymer length,
but also larger than the tube diameter.  For quantification we
consider the free energy cost $\Delta F(\gamma)$ of confining the
polymer to the tube. It can be found from the partition sum that is
obtained as a path integral over all polymer configurations:
\begin{equation} \label{eq:free_energy}
\exp \left[ -\beta \Delta F(\gamma) \right] = \int {\cal D}[{\bf
  r}_\perp(s)] \exp[-\beta H(\kappa,\gamma)]
\end{equation}
with $\beta=1/k_{\rm B} T$.
In the limit of infinitely long polymers the free energy cost is
\cite{burkhardt95}
\begin{equation} \label{eq:free_energy_result}
\Delta F = \sqrt{2} k_{\rm B} T \, \frac{L}{L_{\rm d}} \;.
\end{equation}
This result fits into the picture of the deflection length as measure
for the average distance between successive collisions of the polymer
with its tube. If the typical distance between two collisions is given
by $L_{\rm d}$, the free energy loss results as the sum over all
$L/L_{\rm d}$ points of contact where every collision costs one
$k_{\rm B} T$. The free energy now allows one to derive the tube
diameter as
\begin{equation}
L_\perp^2 = \frac{2}{L} \, \frac{\partial \Delta F}{\partial \gamma}
          = \frac{L_{\rm d}^3}{\sqrt{2}l_{\rm p}} \;.
\end{equation}
In the limit of infinite polymer length we have thus derived the tube
diameter as a function of the deflection length by differentiation of
the free energy cost.

The above consideration also sets the road map for the remaining work.
To calculate the tube diameter for the network, we need first to
connect free energy and tube diameter for polymers of finite length
and then derive the deflection length for the model under
investigation.

\subsection{Finite length Polymers}
For finite size polymers the path integral in
Eq.(\ref{eq:free_energy}) can be evaluated exactly
\cite{burkhardt95,kleinert86} and with the dimensionless deflection
length $l_{\rm d} := L_{\rm d}/L$ results in
\begin{equation}
\Delta F= - 2 k_{\rm B} T g(l_{\rm d})
\end{equation}
with 
\begin{equation} \label{eq:definition_g}
  g(l_{\rm d})=\ln(l_{\rm d}^2)-\frac{1}{2} \ln \left(\sinh^2 \frac{1}{\sqrt{2}
      l_{\rm d}}-\sin^2 \frac{1}{\sqrt{2} l_{\rm d}} \right) \;.
\end{equation}
The limit of small $l_{\rm d}$ that is guaranteed by $L \gg L_{\rm d}$ as
stated above, allows an expansion
\begin{equation} \label{eq:approx_g}
  g(l_{\rm d})=-\frac{1}{\sqrt{2}l_{\rm d}}+\ln(l_{\rm d}^2)
               +{\cal O} (e^{-1/l_{\rm d}}) \quad
  \mathrm{for} \quad l_{\rm d} \to 0 \;,
\end{equation}
where the first term is just the result for polymers with infinite
length (\ref{eq:free_energy_result}).  Upon again using the relation
$L_\perp^2 = (2/L) (\partial \Delta F (l_d)/ \partial \gamma)$ with
the inner derivative $\partial l_{\rm d}/ \partial \gamma = -
(L^4/4\kappa)l_{\rm d}^5$ the tube diameter becomes
\begin{equation}
L_\perp^2=\frac{L^3}{2 l_{\rm p}} l_{\rm d}^5 g^\prime(l_{\rm d}) \;.
\end{equation}
For later convenience we simplify this to $l_\perp^2=h(l_{\rm d})$ by
introducing a dimensionless tube width $l_\perp$ and function $h(x)$
as
\begin{equation} \label{eq:definition_h}
l_\perp^2 := \frac{L_\perp^2 l_{\rm p}}{L^3} \qquad \mathrm{and} \qquad
h(x):=\frac{x^5 g^\prime(x)}{2} \;.
\end{equation}
This relation connects the wanted tube diameter to the deflection
length and hence to the hypothetical tube potential $\gamma$ at a
given bending rigidity. In the following we will further investigate
the tube properties and set up a model that allows one to derive the
deflection length and thereby the hypothetical harmonic tube potential
strength from the polymer concentration and persistence length.

\section{Independent Rod Model}\label{sec:indep-rod-model}

\begin{figure}[htbp]
\center
\includegraphics[width=\columnwidth]{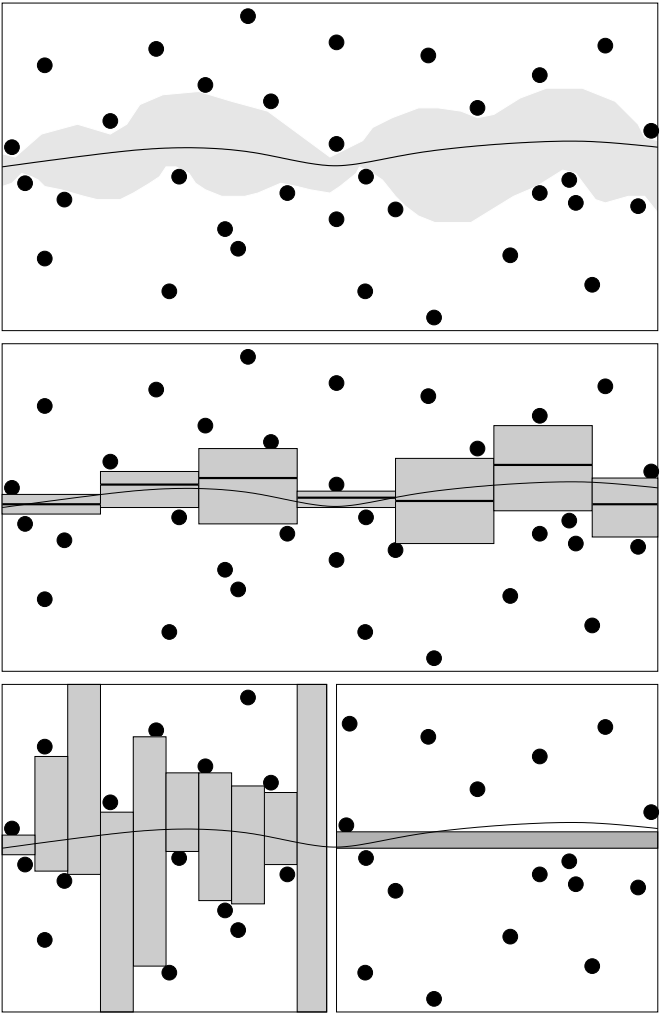}
\caption{(top) The fluctuation tube of a semiflexible polymer in a
  network of constraints is determined by a delicate balance of
  entropic and bending energy. (middle) Scheme of decomposition of a
  semiflexible polymer into rigid rods of length $\bar L$. The
  flexibility is localized to the joints between independent rods.
  Given the proper choice of $\bar L$ both models produce the same
  transverse fluctuation area. (bottom) Small rod length
  overestimates and large rod length underestimates fluctuations.}
\label{fig:irm}
\end{figure}

For simplification and as an anticipation towards the computer
simulations, consider for the time being a polymer in a
two-dimensional (2D) plane. In this case the transverse displacement
vector ${\bf r_\perp}$ reduces to a single component. The undulations
of the test polymer in 2D are hindered by point-like obstacles as
depicted in Fig.~\ref{fig:irm} (top). These obstacles represent the
cuts of the surrounding polymers in three dimensions with the chosen
fluctuation plane. Given an appropriate number of 2D obstacles
equivalent to the density of surrounding polymers in 3D, the
transverse displacement will correspond to one of the two components
of the displacement vector ${\bf r_\perp}$, if we assume the
fluctuations of these components to be independent. Bearing in mind
the large persistence length compared to the mesh size, the
surrounding polymers in 3D are modelled as rigid rods and the area
density $\rho_{\rm MC}$ of obstacles in 2D corresponding to a polymer
concentration $\nu$ in the 3D network is $\rho_{\rm MC}=2 \nu L/\pi$.
It is computed in Appendix \ref{sec:density} and will be explicitly
needed for the comparison with simulation results.

Recalling the Hamiltonian (\ref{eq:hamilton}), the polymer's free
energy has a bending and an entropic contribution. To minimize the
free energy it can be favorable to trade in bending energy for a wider
tube. Thereby entropy is gained due to a larger available free volume,
but the polymer is forced to sacrifice energy to obtain its curvature
(see Fig.~\ref{fig:irm} (top)). This competition defines a
characteristic length $\bar L$ that has to be of the order of the
deflection length $L_{\rm d}$, since this is the length scale
characterizing interaction of the test polymer and its environment.

In the following we will develop an analytical theory based on an
independent rod model (IRM) that is inspired by the competition we
have just discussed. To this end, we use a simplified model of a
semiflexible polymer, in which the flexibility is localized to the
joints of a sequence of independent stiff rods of length $\bar L$.
After deriving the transverse fluctuations of a single independent rod
in an environment of fluctuating neighbors, we apply a
self-consistency argument to arrive at the corresponding tube width of
the full length semiflexible polymer. Note that the analysis is
carried out for three dimensions and the 2D simplification only serves
for illustration and for simulations later on.

To begin with, consider the test polymer to be divided into
independent segments of length $\bar L$ that are assumed to be
completely rigid rods and are only allowed to undergo transverse
fluctuations. As the flexibility in the IRM depends on the number of
joints, it is obvious that the choice of $\bar L$ is crucial for the
resulting tube diameter. Picturing the decomposition of the test
polymer as in Fig.~\ref{fig:irm} (middle) it can be seen that the
transverse fluctuations of the independent rods are hindered by the
two closest obstacle polymers normal to either side of each segment of
length $\bar L$. If $\bar L$ is chosen too large (e.g. $\bar L=L$ in
the worst case) the area of transverse fluctuations will be much
smaller than for a true semiflexible polymer because flexibility is
underestimated (Fig.~\ref{fig:irm} (bottom, right)). On the contrary,
if $\bar L$ is chosen too small, the normal distance to the nearest
obstacle can be quite large (Fig.~\ref{fig:irm} (bottom, left)). This
overestimation of flexibility results in a transverse fluctuation area
that is large compared to the polymer we try to model. Before we
further discuss the proper choice of $\bar L$, we will focus on the
behavior of a single independent rod in more detail.

\begin{figure}[htbp]
\center
\includegraphics[width=\columnwidth]{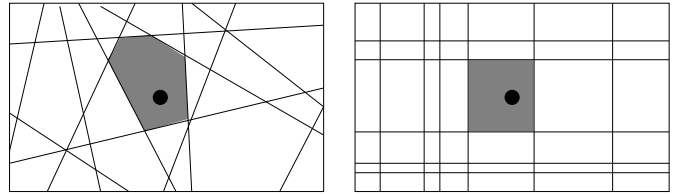}
\caption{(left) Projection of constraining polymers to the plane of
  transverse fluctuations of a test polymer (black dot). As the mesh
  size is much smaller than the persistence length, the constraining
  filaments, can be assumed to be straight. The shaded area is the
  accessible tube area for a specific obstacle configuration. (right)
  Corresponding setup for a simplified geometry where obstacles can
  only be aligned with coordinate axes. }
\label{fig:crossection}
\end{figure}

The transverse fluctuation of a single stiff rod in the $(y,z)$-plane
are constrained by the projections of the surrounding network
constituents to this plane as depicted in Fig.~\ref{fig:crossection}
(left).  Since the mesh size is much smaller than the persistence
length, the surrounding polymers can be assumed to be straight and
``dangling ends'' are neglected. The size of the shaded cross section
will decrease with increasing density of polymers, i.e. with a
decreased mesh size. Thus the tube diameter is of order of the mesh
size and scales as $L_\perp \propto \xi$ for a given 2D plane.
Furthermore, an increase of the length $\bar L$ of the rigid rod
signifies an increase of obstacles that will be projected to the
plane. As the average distance between surrounding polymers in
direction of the test rod is also given by the mesh size $\xi$, the
average number projected onto the plane increases as $\bar L/\xi$. As
this reduces the cross section area, we finally arrive at an overall
scaling of the tube diameter as $L_\perp \propto \xi^2/\bar L$.

Before we quantify this scaling result in the next section, let us
first have a closer look at the obstacles. In a self-consistent
treatment these evidently have themselves to be regarded as
semiflexible polymers of the network and therefore undergo
fluctuations around an average position as well.  This causes the
cross section area to smear out, as the test polymer has now a
non-vanishing probability to take on values behind the average
obstacle position. In terms of a confinement potential the cross
section is no more described by an infinite well, but by some
continuous potential which earlier has been assumed to be harmonic
with strength $\gamma$ per unit length of a polymer. The obstacle
fluctuations will also be modelled as Gaussian and to distinguish
between the test polymer mean square displacement $L_\perp^2$ and the
obstacle's, the latter is denoted as $\sigma^2$. In a self-consistent
treatment of the network the average tube width $L_\perp$ of the test
polymer is then determined as a function of the obstacle fluctuations
$\sigma$, where $\sigma$ is chosen such that $L_\perp=\sigma$.  Of
course, the value $L_\perp$ of a single obstacle configuration will
not only depend on $\sigma$ but also on the obstacle positions in that
specific configuration.  Consequently, averaging over all obstacle
configurations will result in a distribution $P(L_\perp)$ and
self-consistency would then also require a distribution $P(\sigma)$.
However, if we assume these distributions to be reasonably peaked, we
can use their averages as a good approximation. The self-consistency
of distributed tube widths is verified by simulations in Section
\ref{sec:simulations}.

\subsection{Single stiff rod in simplified geometry}

According to the assumptions made above the obstacles (in a top view)
are completely described by a normal distance $r_k$ from the test
polymer and an orientation $\alpha_k$; compare
Fig.~\ref{fig:crossection}. We will neglect correlations and assume the
obstacles to be uniformly distributed. The probability to find an
obstacle with a certain direction at a specified point is independent
of the direction and that point. This corresponds to a complete
factorization of the network distribution function into single polymer
distribution functions.

Consider first a simplified geometry in which all obstacles are either
parallel to the $y$ or the $z$ axis as depicted in
Fig.~\ref{fig:crossection} (right). As fluctuations in both
coordinates are assumed to be independent and equivalent, the task of
computing the tube width is reduced to a one dimensional problem with
a single coordinate $r$. The network density or mesh size enters as
the number $\rho$ of obstacles per unit length. This density should be
chosen such, that the average number of obstacles at a certain
distance $r$ from the test rod in the IRM is the same as the average
number of obstacle polymers featuring a minimal distance $r$ from the
test polymer. This density is proportional to the length $\bar L$ of
the stiff segment and the number of surrounding polymers in a unit
volume $\nu L$.  The exact relation $\rho=(\pi/2) (\nu L \bar
L/4)$ is calculated in Appendix \ref{sec:density2}. 

As the obstacles are assumed to undergo Gaussian fluctuations around
their average position $r_k$, the corresponding probability
density is
\begin{equation}
P_0(r-r_k,\sigma):=\left( 2 \pi \sigma^2  \right)^{-1/2}
e^{\frac{-(r-r_k)^2}{2 \sigma^2}} \;.
\end{equation}
If the test rod interacts with only a single obstacle, we can state
that the probability to find the test rod at a certain position is
given by the fraction of realizations still accessible to the obstacle.
In this case
\begin{equation}
  P_+(r,r_k,\sigma)=\int_r^\infty dr^\prime
  P_0(r^\prime-r_k,\sigma)
\end{equation} 
is the fraction of configuration space still accessible to the obstacle
if the test rod is placed at $r$ (for $r_k>0$). Completing the
integral yields
\begin{equation}
P_+(r,r_k,\sigma) = \frac{1}{2} \mathrm{erfc} 
                    \left( \frac{r-r_k}{\sqrt{2}\sigma} \right)
\end{equation}
and the corresponding probability for obstacles at negative positions
$P_-(r,r_k,\sigma)$ is simply obtained by a inverted sign of the
argument. The probability to find the test rod at a position $r$ for a
given configuration of obstacles $\{r_k\}$ is then given by the
product of all probabilities
\begin{equation} \label{eq:product}
P(r,\{r_k\},\sigma) = \frac{1}{{\it N}}
\prod_{k,r_k>0} P_+(r,r_k,\sigma) \prod_{k,r_k<0} P_-(r,r_k,\sigma) \;.
\end{equation}
The normalization ${\it N}={\it N}(\{r_k\},\sigma)$ is determined by
the condition $\int dr P(r,\{r_k\},\sigma)=1$ and depends on the
obstacle configuration. 

\begin{figure}[htbp]
\center
\includegraphics[width=\columnwidth]{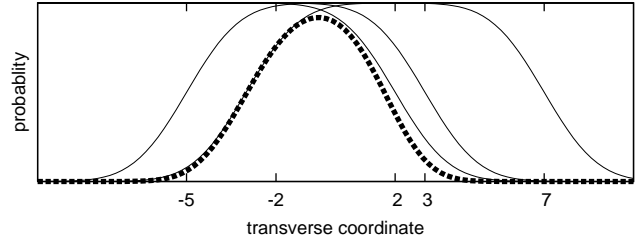}
\caption{Probability density to find the test rod at a spatial 
  position for mutual interaction with a single obstacle (solid lines)
  and resulting probability in an environment of all obstacles (dashed
  line). The x-axis tics mark the center position of each obstacle.
  Distant obstacle only have a negligible influence on the overall
  probability function. }
\label{fig:probability}
\end{figure}

As the function $P_+(r,r_k,\sigma)$ reduces to a Heaviside function in
the case of $\sigma \to 0$, the product in Eq. (\ref{eq:product}) can
be written as $\theta(r-r_-)\theta(r-r_+)/(r_+-r_-)$ where $r_+$ and
$r_-$ are the positions of the two closest obstacles. This reduction
is justified because all further obstacles are completely shadowed by
the two nearest neighbors. In the case of a non-vanishing $\sigma$ the
probability distribution $P(r,\{r_k,\alpha_k\},\sigma)$ will not be
rectangular anymore but smear out. The test rod has a non-vanishing
probability to be found behind the average position of the closest
obstacle and thus a chance to feel the interaction of further network
constituents. However, sketching the distribution in
Fig.~\ref{fig:probability}, it becomes intuitively clear that this
probability rapidly approaches zero for far obstacles or small
fluctuation amplitudes $\sigma$. We will exploit this fact in the
numerical analysis below and in the
simulations.

The distribution function, Eq.~(\ref{eq:product}), for the test rod at
hand, averages of any function  $f(r)$ can now be calculated for a single
realization of obstacles as
\begin{equation} \label{eq:test_average}
\overline{f(r)}_{\{r_k\}}=\int dr f(r) P(r,\{r_k\},\sigma) \;,
\end{equation}
where the index $\{r_k\}$ denotes the specific obstacle configuration.
The tube center of the test rod is then
\begin{equation}
\overline{r} (\{r_k\},\sigma):= \overline{r}_{\{r_k\}}
\end{equation}
and the width of the probability distribution is the wanted tube
diameter
\begin{equation}
  L_\perp^2(\{r_k\},\sigma) 
 := \overline{r^2}_{\{r_k\}}-\overline{r}^2_{\{r_k\}} \;.
\end{equation}
The derived tube diameter of the test rod is not only a function of
the fluctuation width $\sigma$ but also of the specific obstacle
configuration. Consequently, sampling over different obstacle sets
will result in a distribution of values for $L_\perp$. As mentioned
earlier this distribution should be described by a single
characteristic value - consistent with the obstacle fluctuations that
have also be assumed to be of equal size. Since the obstacles are
uniformly distributed, they can be fully described by the density
encoded in the average number of obstacles $\rho$ per line. This is
achieved by integrating out all obstacle positions and orientations in
$L_\perp^2(\{r_k\},\sigma)$ to arrive at the only density dependent
$L_\perp^2(\rho,\sigma)$. We choose a simple average over a large
number $N$ of obstacle sets $\{r_k\}$ like
\begin{equation} \label{eq:obstacle_integration}
  \langle f(\{r_k\}) \rangle_\rho = \left( \prod_{k=1}^N
    \int_{-R/2}^{R/2} \frac{dr_k}{R}\right) f(\{r_k\}) \;,
\end{equation}
where $R=N/\rho$. In this nomenclature the average tube
diameter~\footnote{Of course, one could also image a different
  characterization of the average tube diameter, e.g. the median or
  the maximal diameter. We choose the average as the most obvious
  quantity experimental groups might measure, e.g.  in analyzing
  different fluorescent microscopy images.} is obtained as
$L_\perp^2(\rho,\sigma)=\langle L_\perp^2(\{r_k\},\sigma)
\rangle_\rho$.

Self-consistency is now expressed as
\begin{equation}
L_\perp^2(\rho,\sigma)=\sigma^2
\end{equation}
at the point of self-consistency (PSC) $\sigma=\sigma^*$. By measuring
length in $1/\rho$ we can rewrite this to a dimensionless
master curve $l(\rho \sigma)$: 
\begin{equation}
L_\perp^2(\rho,\sigma)=\frac{1}{\rho^2} l(\rho \sigma) \;,
\end{equation}
since $L_\perp, 1/\rho$ and $\sigma$ are all lengths, The task of
finding the self-consistent tube width
$L_\perp(\rho,\sigma^*)=\sigma^*$ translates to finding $l(C)=C^2$
where the constant $C=\rho \sigma^*$. As soon as this is achieved, the
self-consistent tube diameter is available as a function of density
$\rho$ only and hence it depends like
\begin{equation} \label{eq:tube_diameter_of_rho}
L_\perp=\frac{C}{\rho}=\frac{2 C}{3 \pi} \frac{\xi^2}{\bar L}
\end{equation}
on the rod length $\bar L$ and mesh size $\xi$.

The numerical determination of $C$ is achieved by an integration using
a Monte-Carlo procedure. It includes the $N$-fold integrals over the
obstacle positions $r_k$ from Eq. (\ref{eq:product}) as well as the
integration over the test polymer position $r$ from Eq.
(\ref{eq:test_average}). As mentioned above, the probability
distribution rapidly decreases at distances far from the closest
obstacle. Hence, we have restricted the integration range of the $dr$
integration in the Monte-Carlo samples to values $[y_{min}-5
\sigma,y_{max}+5 \sigma]$ where $y_{min}$ and $y_{max}$ are the
closest obstacle at either side. Furthermore, the fast decrease of the
probability distribution renders the contribution of distant obstacles
quasi to zero. We can therefore drop all obstacles with $y_k \not\in
[y_{min}-10 \sigma, y_{max}+10 \sigma]$.  The results are depicted in
Fig.~\ref{fig:mciso} and a graphical solution for the PSC constant
results in
\begin{equation}
C \approx 3.64 \;.
\end{equation}
Special attention should be paid to the behavior at $\rho \sigma=0$.
It provides a good test whether the used IRM is adequate and allows
for a verification of the numerics. At a finite density as required by
the tube concept, $l(0)$ reflects the situation of immobile obstacles
with $\sigma=0$. At this point the tube diameter should remain finite
and its value should be given by the density of obstacles. From the
obstacle statistics and density per unit length $\rho/4$, the
probability to find the first obstacle at position $r_\pm$ is known to
be $P(r_\pm) = \exp(-r_\pm \rho/4)$. In the case of fixed obstacles
the available fluctuation area is $2 L_\perp=r_+-r_-$ and the
expectation value $4 \langle L_\perp^2 \rangle=\langle (r_+-r_-)^2
\rangle$ can be computed from the probability density above.  Taking
care of the normalization one arrives at $L_\perp=\sqrt{8}/\rho$. The
master function yields $l(0)=\rho^2 L_\perp^2=8$, a value in good
agreement with the data (circles) in Fig.~\ref{fig:mciso}.

\subsection{Generic 2d Geometry}

If the simplification of axis-parallel obstacle polymers is dropped
again, the obstacle configuration needs to be specified by a set of radii
$\{r_k\}$ and angles $\{\alpha_k\}$. The
probability to find the test rod at a position $(y,z)$ for a given
configuration of obstacles $\{r_k,\alpha_k\}$ in the two-dimensional
case as in Fig.~\ref{fig:crossection} (left) is then again given by
the product of all probabilities where different angles have to be
accounted for:
\begin{equation} \label{eq:product_iso}
P(y,z,\{r_k,\alpha_k\},\sigma)=\frac{1}{{\it N}}
\prod_k P_{\pm}(y \cos \alpha_k + z \sin \alpha_k, r_k, \sigma) \;.
\end{equation}
The normalization factor ${\it N} = {\it N}(\{r_k,\alpha_k\},\sigma)$
is again determined by the condition $\int dy dz
P(y,z,\{r_k,\alpha_k\},\sigma)=1$.

\begin{figure}[htbp]
\center
\includegraphics[width=\columnwidth]{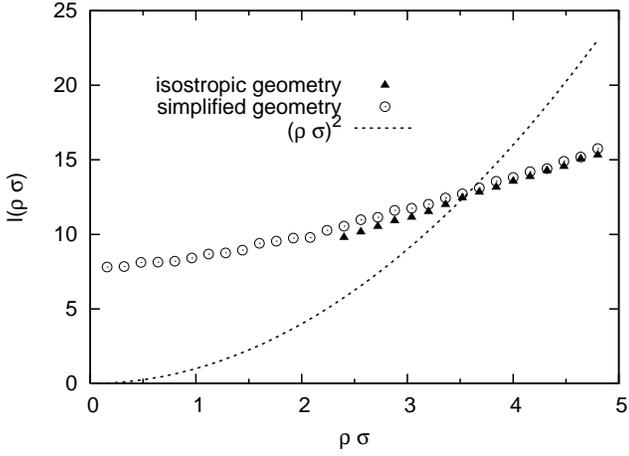}
\caption{Master curve $l(\rho \sigma)$ of the tube diameter rescaled
by obstacle density obtained by MC simulation for simplified (circles)
and generic geometry (triangles); intersection with quadratic obstacle
fluctuation amplitude marks the point of self-consistence. The error of
the simplified geometry is surprisingly small.}
\label{fig:mciso}
\end{figure}

In a single obstacle configuration the tube diameters $L_{\perp y,z}$
in the $y$ and $z$ direction will in general be different. However, in
averaging over all configurations isotropy must be recovered to show
\begin{equation}
  L_\perp^2(\rho,\sigma)=\langle
  L_{\perp y}^2(\{r_k,\alpha_k\},\sigma) \rangle_\rho=\langle
  L_{\perp z}^2(\{r_k,\alpha_k\},\sigma) \rangle_\rho \;.
\end{equation}
The average over obstacle configurations at fixed density of uniformly
distributed obstacles is performed as
\begin{equation} \label{eq:iso_obstacle_integration}
  \langle f(\{r_k,\alpha_k\}) \rangle_\rho = \left( \prod_{k=1}^N
    \int_0^R \frac{dr_k}{R} \int_0^{2 \pi} \frac{d\alpha_k}{2 \pi }\right) f(\{r_k,\alpha_k\})
\end{equation}
with integration range being again $R=N/\rho$. Note that contrary to
the simplified geometry the obstacle density per unit length $\rho$ in
this case is given by $\rho=(\pi/2) (\nu L \bar L)$.  Evaluating the
integrals in Eq.  (\ref{eq:iso_obstacle_integration}) again by the
Monte-Carlo method results in the data plotted in Fig.~\ref{fig:mciso}
(triangles), where suppression of irrelevant obstacles was implied
analog to the simplified geometry.
The results do not deviate much from the data obtained earlier
(circles), i.e. the mistake in using a simplified geometry is
surprisingly small. Again the value of the PSC is obtained
graphically. It yields
\begin{equation}
C \approx 3.52
\end{equation}
and will be used in the remainder of this work.

\subsection{Choice of Independent Rod Length}
As discussed before and illustrated in Fig.~\ref{fig:irm} (bottom) the
choice of $\bar L$ is crucial for the success of the IRM. The number
$L/\bar L$ of independent rods can be regarded as a measure for the
flexibility of the modeled polymer and has to be chosen such that the
transverse excursions of the ensemble of stiff rods equal the
fluctuations of the actual semiflexible polymer. To this end we
consider both systems in a generic harmonic potential
\begin{equation}
U[y(s)]=\frac{\gamma}{2}\left[y(s)-y^0(s) \right]^2
\end{equation} 
with the potential minimum $y^0(s)$ as a Gaussian variable with
$\langle y^0(s) y^0(s') \rangle = \alpha \delta(s-s')$. This
corresponds to the assumption of a ``Gaussian random backbone'' as a
general property of the tube. We use this intuitive assumption as one
possible prerequisite to determine the segment length $\bar L$. Of
course, other possibilities can be imagined. Note that the simulations
in Sec. \ref{sec:simulations} will justify this assumptions a
posteriori.

The average position $\overline{y(s)}$ as a functional of a given
potential $y^0(s)$ is obtained as an average over all polymer
configurations in this potential. Averaging then over all potential
conformations yields the the mean square of the polymer's transverse
fluctuations $\langle \overline{y(s)}^2 \rangle$. The over-line thus
denotes an average in a given potential and the brackets denote an
average over all potentials. While the transverse fluctuations of a
rigid rod are a function of the potential parameters $\alpha, \gamma$
only, the response of a semiflexible polymer will additionally depend
on its stiffness. This evidently provides a tool to connect the
semiflexible polymer persistence length and the length $\bar L$ from
the IRM by demanding that the fluctuations $\langle \overline{y(s)}^2
\rangle$ for given potential parameters $\alpha, \gamma$ are the same
for both cases.

Starting with the IRM, it is sufficient to consider only one stiff
rod, as the individual rods are statistically independent. The average
position is then
\begin{equation}
\overline{y}=\frac{1}{\bar L} \int_0^{\bar L} \textrm{ds} y^0(s)
\end{equation}
and the transverse fluctuations
\begin{equation}
\langle \overline{y}^2 \rangle = \frac{1}{\bar L^2} \int_0^{\bar L}
ds \int_0^{\bar L} ds' \langle y^0(s) y^0(s') \rangle =
\frac{\alpha}{\bar L} \;.
\end{equation}
For the semiflexible polymer the fluctuations of polymer and tube
potential are decomposed into modes (Appendix
\ref{sec:mode-analys-polym}):
\begin{equation}
\langle \overline{y}^2 \rangle = \frac{1}{L} \int_0^L \langle
\overline{y(s)}^2 \rangle = \frac{1}{L} \sum_k \langle
\overline{y_k}^2 \rangle \;,
\end{equation}
where the mode analysis yields $\overline{y_k}=y_k^0/(1+q_k^4 l_{\rm d}^4)$
(compare Eq.~(\ref{eq:app2_1})) with $q_k \approx \pi (k+1/2)$. Using
now the correlations of the Gaussian random tube profile and the
identity (\ref{eq:app2_3}) the polymer fluctuations can be related to
the deflection length as:
\begin{equation}
\langle \overline{y}^2 \rangle=\frac{1}{L} \sum_k \frac{\langle
  (y_k^0)^2 \rangle}{(1+q_k^4
  l_{\rm d}^4)^2}=\frac{\alpha}{L}\frac{h'(l_{\rm d})}{4 l_{\rm d}^3} \;.
\end{equation}
Equating the fluctuations for the IRM and the semiflexible polymer
fixes the segment length to
\begin{equation} \label{eq:rod_length}
\bar L=L \frac{4 l_{\rm d}^3}{h'(l_{\rm d})} \;.
\end{equation}
Concluding the last section, we have obtained the tube diameter for a
sequence of independent rods of length $\bar L$ and derived a
condition how to fix this length to correctly mimic the behavior of
a semiflexible polymer in a network of same mesh size. It has turned
out the the criteria for the correct rod size is a function of the
deflection length.

\section{Results}\label{sec:plugging-it-all}

If we recall that the tube diameter for a semiflexible polymer was
derived in Sec. \ref{sec:model-definition} from the Hamiltonian with a
likewise dependence on deflection length, we are now equipped to set
up an implicit equation to determine this deflection length.
Afterwards the tube diameter can be derived from a simple
calculation. 

Equating the expressions for the tube diameter of the polymer
(\ref{eq:definition_h}) and the IRM (\ref{eq:tube_diameter_of_rho})
respectively yields
\begin{equation}
L_\perp^2=\frac{L^3}{l_{\rm p}} h(l_{\rm d})=\frac{4 C^2}{9 \pi^2}
\frac{\xi^4}{\bar L^2} \;.
\end{equation}
With the correct rod length (\ref{eq:rod_length}) the implicit
equation for the dimensionless deflection length is 
\begin{equation} \label{eq:implicit_of_h}
h(l_{\rm d}) = \frac{C^2}{32 \pi^2} 
               \frac{[h'(l_{\rm d})]^2}{l_{\rm d}^6} 
               \frac{l_{\rm p} \xi^4}{L^5} \;.
\end{equation}
Solving this equation, determines $l_{\rm d}$ from the system's parameter
$l_{\rm p}, L$ and $\xi$. It is achieved by introducing a dimensionless
function
\begin{equation}
l_{\rm p} \xi^4/L^5=j(l_{\rm d})
:=\frac{l_{\rm d}^6 32 \pi^2 h(l_{\rm d})}{C^2 [h'(l_{\rm d})]^2 } \;.
\end{equation}
Inversion then yields
\begin{equation}
l_{\rm d}=j^{-1}(l_{\rm p} \xi^4/L^5) \;.
\end{equation}
With the abbreviation $D=3 C/\pi$ we finally obtain for the
dimensional deflection length to first order and second order in the
argument of $j^{-1}$:
\begin{equation}
L_{\rm d}=\frac{D^{2/5}}{2^{13/10}} \xi^{4/5} l_{\rm p}^{1/5} +
\frac{D^{4/5}}{2^{11/10}3}\frac{\xi^{8/5}l_{\rm p}^{2/5}}{L} \;.
\end{equation}
By application of (\ref{eq:definition_h}), the tube diameter is easily
obtained as
\begin{equation} \label{eq:tube2order}
L_\perp=\frac{D^{3/5}}{2^{27/10}}\frac{\xi^{6/5}}{l_{\rm p}^{1/5}} +
\frac{D}{2^{5/2}} \frac{\xi^2}{L} \;.
\end{equation}
Evaluation of the numerical factors holds the following results to
first order
\begin{equation}
L_{\rm d} \approx 0.66 \xi^{4/5} l_{\rm p}^{1/5} \;, 
\qquad L_\perp \approx 0.32
\frac{\xi^{6/5}}{l_{\rm p}^{1/5}} \;.
\end{equation}
Note that this also determines the confinement free energy of the
polymer to $\Delta F \approx 2.14 \, k_{\rm B} T \, \frac{L}{\xi^{4/5}
  l_{\rm p}^{1/5}}$.  The leading term of the tube diameter agrees
with the established scaling \cite{semenov86}. The additional term's
dependence on the inverse polymer length indicates a finite length
effect. It can be traced back to the partition sum of a finite polymer
(\ref{eq:approx_g}) and accounts for boundary effects at the end of
the tube. If the free energy of infinite polymers
(\ref{eq:free_energy_result}) is used throughout the calculations, all
higher order terms vanish accordingly. In an earlier work
\cite{morse01} another prefactor of $L_\perp \approx 0.53 \xi^{6/5}
l_{\rm p}^{-1/5}$ for the scaling term has been predicted by rather
different accounting of obstacles.

\begin{figure}[htbp]
\center
\includegraphics[width=\columnwidth]{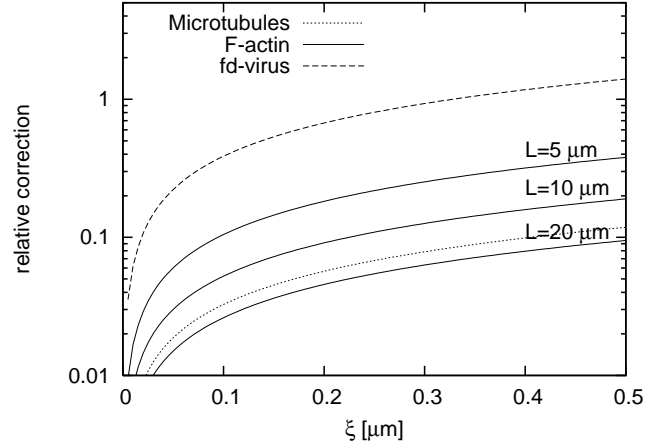}
\caption{Relative correction obtained by the second order term of the
  tube diameter (\ref{eq:tube2order}) for different biopolymers as a
  function of mesh size $\xi$. fd-viruses ($L \approx 0.9 \mu$m, $l_{\rm p}
  \approx 2.2 \mu$m \cite{schmidt00}) show a large correction due to
  their small length compared to the mesh size, while this effect is
  rather small in microtubules ($L \approx 50 \mu$m, $l_{\rm p} \approx 5000
  \mu$m \cite{pampaloni06,gittes93}). The correction for F-actin has
  been plotted for different length from a typical length
  distribution.}
\label{fig:dev}
\end{figure}

It is important to be aware of the subtle difference between the
explicit length dependence of the first order term and the implicit
dependence on $L$ that enters via the mesh size $\xi=\sqrt{3/\nu L}$.
In a polydisperse polymer solution the $L$ in the mesh size has to be
the average polymer length, while the $L$ in the second order terms is
the length of the actually observed filament in the tube. In a
monodisperse solution as in our theory these quantities are identical.

The importance of the second order term depends heavily on the nature
of the polymers making up the network. In Fig.~\ref{fig:dev} the
relative tube width correction obtained by the second order term is
displayed for several semiflexible biopolymers as a function of mesh
size $\xi$. It is interesting to note that the intuitive dependence on
the relative persistence length $l_{\rm p}/L$ present in the second
order term of the deflection length is rather negligible. The most
dominant effect of the correction term is not obtained for the
stiffest biopolymer, a microtubule, but for the small fd-virus. This is
due to its small length to mesh size ratio. Finite length effects will
influence a large fraction of the polymer strand and not only the
boundaries.  Given a proper control of polymer length, this effect
should be experimentally observable in F-actin solutions.

\begin{figure}[htbp]
\center
\includegraphics[width=\columnwidth]{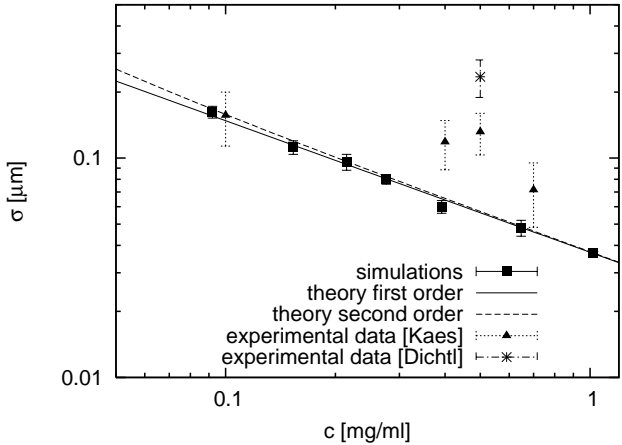}
\caption{Comparison of tube diameter from theory, numerical
  simulations (squares) and reanalyzed experimental measurements
  (triangles) from \cite{kas96,dichtl99}. While Dichtl has directly
  measured potential strengths, K\"as has recorded the maximal tube
  width $a$.  Therefrom we estimated a lower boundary of
  $\sigma=a/6$.}
\label{fig:results}
\end{figure}

Focussing back on F-actin, Fig.~\ref{fig:results} displays the result
of our model in comparison to experimental data \cite{kas96,dichtl99}.
While theoretical and experimental results are certainly qualitatively
comparable, a more detailed discussion is difficult due to the large
fluctuations of the measurements. However, it seems reasonable to
interpret these measurements regardless of their ambiguity as an upper
limit to the tube diameter. Two main reasons cause an experimental
observation of tube widths systematically higher than in the presented
theory: from a technical point of view the microscope resolution
broadens the observed tubes.  Additionally, this effect is further
enhanced by collective fluctuations of the complete elastic medium
that remain unaccounted for in our approach. Contrary, the computer
simulations presented below, can be tailored to avoid these effects
and study the exact model system used by the theory.

\section{Simulations} \label{sec:simulations} 

We have conducted intensive numerical simulations of the model system
for several reasons: on the one hand they serve as a tool to verify
the validity of several approximations used in the theoretical
description developed above, being for example the harmonic
description of the tube potential or the assumption of a single
fluctuation amplitude for the obstacles.  Furthermore, the comparison
between the simulated transverse fluctuations and the final result of
our theory can prove if we have succeeded in correctly predicting the
tube diameter in a network of semiflexible polymers. Finally, the
simulations give us the chance to analyze observables that go beyond
the analytical theory presented. These are in particular distribution
functions and open up a further possibility to comparison with
experiments.

We use a Monte Carlo simulation of a single polymer in two dimensions
that is surrounded by point-like obstacles. This reduction will result
in an equal fluctuation amplitude as in the 3D model, because we have assumed
the fluctuations along the different coordinates to be independent.
Simulating a test polymer in 2D and measuring its transverse
displacement, will thus on average correspond to either $L_{\perp,y}$
or $L_{\perp,z}$ given that the number of obstacle points has been
chosen correctly. We calculate this number as the number of stiff rods
that cut an arbitrary unit area plane if these rods are of length $L$,
density per unit volume $\nu$ and equally distributed both in position
and orientation. The approximation as rigid rods is justified by the
large persistence length compared to the mesh size. The relation
between polymer concentration $\nu$ and point-density in simulations
$\rho_{\rm MC}$ then yields (\ref{eq:app1_1}):
\begin{equation}
\rho_{\rm MC}=\frac{2}{\pi} \nu L \;.
\end{equation}
Of course, the obstacles will cut the plane under different angles.
These can be incorporated via different statistics of the obstacle
fluctuations. However, simulations show that no significant
differences compared with orthogonal cuts occur. This can be explained
by an averaging out of anisotropies in performing ensemble averages.
We therefore choose to assume orthogonal intersections of the obstacle
polymers with the plane of simulation.

Having defined a suitable conversion for the polymer density, we will
proceed to implement the other contribution to the test polymer's
Hamiltonian in our simulations, i.e. the persistence length. The test
polymer is modelled as a chain of $N$ rigid segments that approximate
the continuous contour of a worm-like chain. The joint angle between
two segments gives rise to a bending energy summed over all bonds:
\begin{equation}
\beta H(\{t_i\})=k \sum_{i=1}^{N-1} {\bf t}_i {\bf t}_{i+1} \;,
\end{equation}   
where the ${\bf t_i}$ are the tangents and k is chosen such to
reproduce the energy of a semiflexible polymer of persistence length
$l_{\rm p}$. The relation in two dimensions is computed to
\begin{equation}
\frac{L}{l_{\rm p}}=-N \ln \left[ \frac{I_1(k)}{I_0(k)} \right]
\end{equation}
with $I_0$ as modified Bessel function of first kind and its
derivative $I_1$.

The simulations start from an equilibrium conformation of the test
polymer and with obstacle centers $r_i^0$ that are uniformly
distributed.  At this point, obstacles are discriminated into those in
the half-spaces left and right of the test polymer. During the
following evolution of the system, every move of an obstacle and every
conformation change of the test polymer is rejected if it would result
in a reclassification of any obstacle into the other half-space.
Besides this constraint the evolution is only governed by the bending
energy of the test polymer and a harmonic potential $U(r_i)=\sigma/2 \,
(r_i-r_i^0)^2$ for every obstacle. During the evolution the transverse
displacements $L_\perp$ of every bond from the average contour are
recorded over the whole evolution. To avoid boundary phenomena, this
is only done in the bulk. The whole procedure is then carried out
repeatedly for different initial sets of random obstacles and random
test polymer conformations. If the computation is repeated for
different values of $\sigma$, a function $L_\perp(\sigma)$ is obtained
from which the point of self consistence $L_\perp(\sigma)=\sigma$ and
its error can be deduced graphically.  Repeating the procedure for
different parameters, holds results for the tube diameter in
dependence of persistence length $l_{\rm p}$ and concentration $\nu L$
and can be compared to the theoretical prediction and the available
scarce experimental data. As displayed in Fig.~\ref{fig:results} the
simulation results and the theoretical prediction to both first and
second order agree remarkably well. On the basis of the available data
any discrimination between first and second order would be bold.
However, it has to be considered that any deviations due to lack of
simulation time or shortcomings in the Monte Carlo moves will tend to
reduce the observed tube width. The obtained simulation results are
thus a lower boundary to the real tube diameter.

Even if the good agreement between the theoretical predicted tube
diameter and the values observed in numerical simulations suggests our
theoretical description to be valid, we employ the developed
algorithms to explicitly check on some of the assumption made in the
course of deriving the tube diameter.

\begin{figure}[htbp]
\center
\includegraphics[width=\columnwidth]{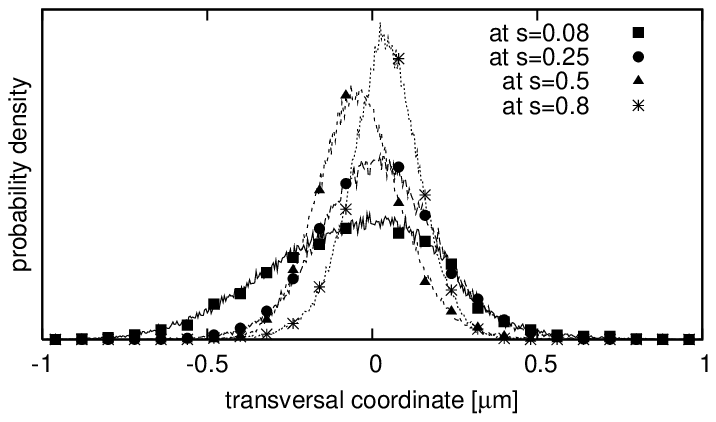}
\hspace{0.5cm}
\includegraphics[width=\columnwidth]{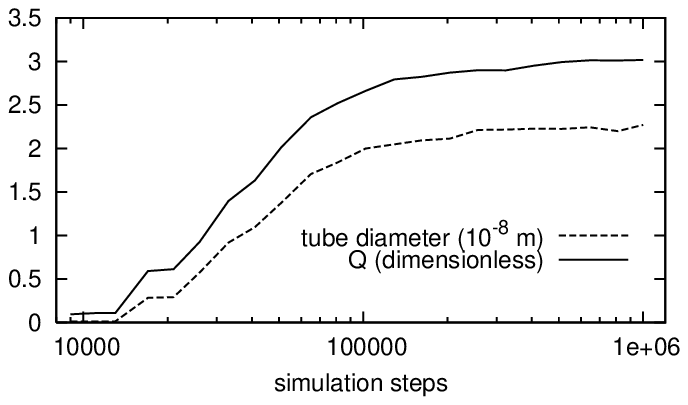}
\caption{(top) Distribution of transverse excursions at different
  arc-lengths $s$ shows a Gaussian potential profile with rather large
  variability in the potential width. At the test polymer's boundaries
  deviations occur. (bottom) After sufficient simulation time the
  ratio $Q$ (solid) approaches the characteristic value $Q=3$ of a
  Gaussian distribution. The transverse fluctuation area (dashed)
  converges likewise.}
\label{fig:harmonic}
\end{figure}

One central assumption in the realm of the tube model is the
substitution of an ensemble of neighboring polymers by an effective
tube potential. This tube potential is modelled by an harmonic
function of strength $\gamma$ as in the Hamiltonian
(\ref{eq:hamilton}). This harmonic assumption seems sensible and is
also supported by preliminary experiments with colloidal probes
\cite{dichtl99}. Our numerical simulations can provide further proof
to the exact form of the potential. To this end, we have monitored the
transverse displacement as a function of arc length. In the resulting
histogram - see Fig.~\ref{fig:harmonic} (top) for some examples - we
identify the distributions maximum as the center position and analyze
the form of the potential. Evidently, the resulting profiles in the
test polymer's bulk are reasonably Gaussian shaped, while deviations
at the boundaries (compare data for $s=0.08$ in
Fig.~\ref{fig:harmonic} (top)) occur but are negligible for a tube
model where $L>L_\perp$. For a quantification the ratio of fourth
moment to square of second moment of transverse fluctuations
\begin{equation}
Q = \frac{\overline{(y-\bar y)^4}}{\overline{(y-\bar y)^2}^2}
\end{equation}
was considered. For a perfect Gaussian distribution this quantity
evolves to $Q=3$. As shown in Fig.~\ref{fig:harmonic} (bottom) this
value is also asymptotically obtained in the simulations after
sufficient simulation time. These results clearly support the validity
of a harmonic tube potential.

\begin{figure}[htbp]
\center
\includegraphics[width=\columnwidth]{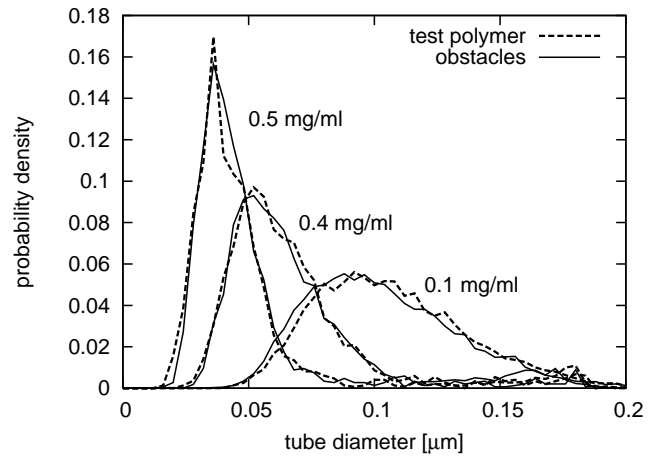}
\caption{Distribution of $L_\perp$ sampled over polymer arc-length and
  different obstacle environments. Distributions are well-peaked and
  exhibit longer tails at large tube widths. The noise at diameters
  far from the distributions maximum is an artefact from the numerical
  discretization.}
\label{fig:distribution}
\end{figure}

In contrast to the classical picture of an Edwards-tube with a
rather homogeneous diameter the simulations reveal a rather large
variability in the local tube diameter as has also been observed
experimentally \cite{kas96,dichtl99}.  Carrying out extensive
simulations in a large number of different obstacle environments
allows one to record the distribution function of the tube diameters.
This is of crucial importance, as our theoretical description has
assumed that the tube diameter - and hence due to self-consistency
also the obstacle fluctuation width - can be described by a single
characteristic value.  This approach only seems feasible if the
distribution described by the characteristic value is reasonably well
peaked. The simulations prove that the resulting distribution is
indeed equipped with a well-defined peak
(Fig.~\ref{fig:distribution}).  However, the variability of the
observed tube diameters is rather large with a half-width of the size
of the average tube diameter itself. We observe a sharp cut-off for
small tube widths while the distribution's tail to wide tubes is
longer. The behavior at small tube width is dominated by the energy
cost of confining a polymer into an increasingly smaller tube and can
thus be considered as a polymer property. On the contrary the
distribution at tube widths larger than the average diameter is due to
void spaces. These will follow an exponential decay and are therefore
a characteristic of the network architecture.

Finally, the numerical simulations provide a means to explicitly check
if the self-consistence is guaranteed in spite of the simplifying
assumption of a single fluctuation width. To this end, we have used
the resulting histogram of tube diameters from above to compute a
normalized distribution function. The fluctuation width of the
obstacles are now initialized according to this very distribution. The
resulting histogram of tube diameters is then again fed back into the
simulation as obstacle fluctuation distribution. This procedure is
carried out until both distributions converge against each other in a
self-consistent manner. Surprisingly, this is already the case after
the first iteration step of the process as displayed in
Fig.~\ref{fig:distribution}. This gives strong evidence that due to
the self-averaging over obstacles the modelling of a network with
Gaussian tube profile and a single average tube diameter is sufficient
to describe the physical reality.

\section{Conclusion}\label{sec:conclusion}

We have presented a new approach to determine the absolute value of
the tube diameter in semiflexible polymer networks supported by
computer simulations. To this end the deflection length of a polymer
in a hypothetical harmonic tube was connected to the tube's diameter
via the free energy cost for finite length polymers. The assumption of
a harmonic tube was confirmed by simulation results. By decomposition
into independent stiff rods of appropriate length, we were able to
establish an implicit equation for the deflection length. The
resulting tube width $L_\perp$ is in agreement with the established
scaling law $L_\perp= c \xi^{6/5}l_{\rm p}^{-1/5}$ with mesh size
$\xi=\sqrt{3/\nu L}$ and persistence length $l_{\rm p}$. Our theory provides
a prefactor of $c \approx 0.32$ and a higher order term that accounts
for finite length effects and scales with $\xi^2/L$.

The available experimental data is consistent with our predictions.
However, its quality does not allow for detailed comparison. To
provide a precise validation, we have complemented our theoretical
work by extensive Monte Carlo simulations of a test polymer in an
environment of obstacles. The resulting self-consistent tube widths
perfectly match the theoretical value predicted. This strongly
supports the validity of the absolute value for the concentration
dependent tube diameter.

Furthermore, we have employed simulations to observe properties
beyond the analytical theory. We have recorded the distribution
function of tube widths in a network for different concentrations.
Thereby we were able to explicitly confirm self-consistency of the
simplifying model with a fixed tube diameter.

Both our theoretical predictions, e.g. the finite length contributions
to the tube diameter, and our simulation data, e.g. the distribution
functions, provide the opportunity of feasible experimental
comparison. On the theoretical side, the significance of correlations
and collective fluctuations of the complete medium, as well as an
analytical model of distribution functions may open up promising
continuations of this work.

\begin{acknowledgement}
  We kindly acknowledge helpful discussions with M.  Degawa, M.
  Giesen, R. Merkel and M. Romanoska.  Financial support of the German
  Excellence Initiative via the program "Nanosystems Initiative Munich
  (NIM)" is gratefully acknowledged. HH acknowledges support by the
  international graduate program Nano-Bio-Technology funded by the
  Elite Network of Bavaria.
\end{acknowledgement}  

\appendix

\section{Rigid Rod Statistics I} \label{sec:density}

To relate the polymer concentration of a network to the obstacle
density per unit area in the simulation, we calculate the number of
randomly distributed and oriented stiff rods per unit volume that
intersect with a unit plane. Every intersecting rod will be described
by its polar and azimuth angle relative to the unit plane, the point of
intersection and the distance between center of mass and intersection
point. Because of rotational symmetry the problem is independent of
the azimuth angle and because of uniform density it is independent of
one of the coordinates of the intersection point. Hence, the problem
is equivalent in two dimensions to the number of rods per unit area
that intersect a unit line (see Fig.~\ref{fig:statistics} (left)). The
plane contains $\nu$ rigid rods per unit area with random orientation
$\alpha$ and center of mass position. The number of intersections
$\rho_{\rm MC}$ with the unit line (bold dashed) is computed by
parameterizing the center of mass (C) by the coordinate $z$ of the
intersection point (P) with the unit line, the distance $s$ between C
and P and the angle $\alpha$. $\rho_{\rm MC}$ is then obtained as the
integral over all possibly intersecting rods:
\begin{equation}
\rho_{\rm MC}=\frac{2 \nu}{2 \pi} \int dr^2 \int_0^{\pi} d\alpha \;,
\end{equation}
where the factor $2$ accounts for the fact that any rod configuration
can be realized by two angles $\alpha$ since the rods have no
direction. The integration area has to be chosen appropriately to only
include intersecting rods. As $r=(\sin(\alpha)s, x-\cos(\alpha) s)$
the Jacobian determinant of the coordinate transformation to
integration variables is $\partial r/\partial(x,s)=\sin(\alpha)$ and
the integral evolves to
\begin{equation} \label{eq:app1_1}
\rho_{\rm MC}=\frac{\nu}{\pi} \int_{-1/2}^{1/2}dx \int_{-L/2}^{L/2} ds
\int_{0}^{\pi} \sin(\alpha)=\frac{2}{\pi}\nu L \;.
\end{equation}

\section{Rigid Rod Statistics II} \label{sec:density2} 
\begin{figure}[htbp]
\center
\includegraphics[width=3.5cm]{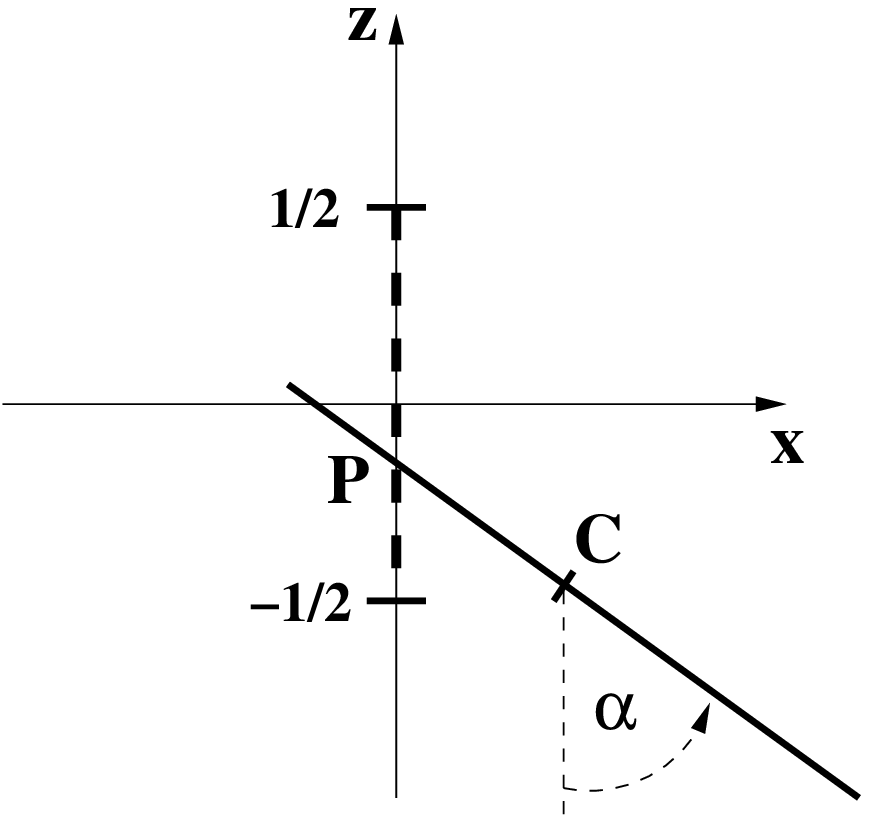}
\hspace{0.5cm}
\includegraphics[width=4cm]{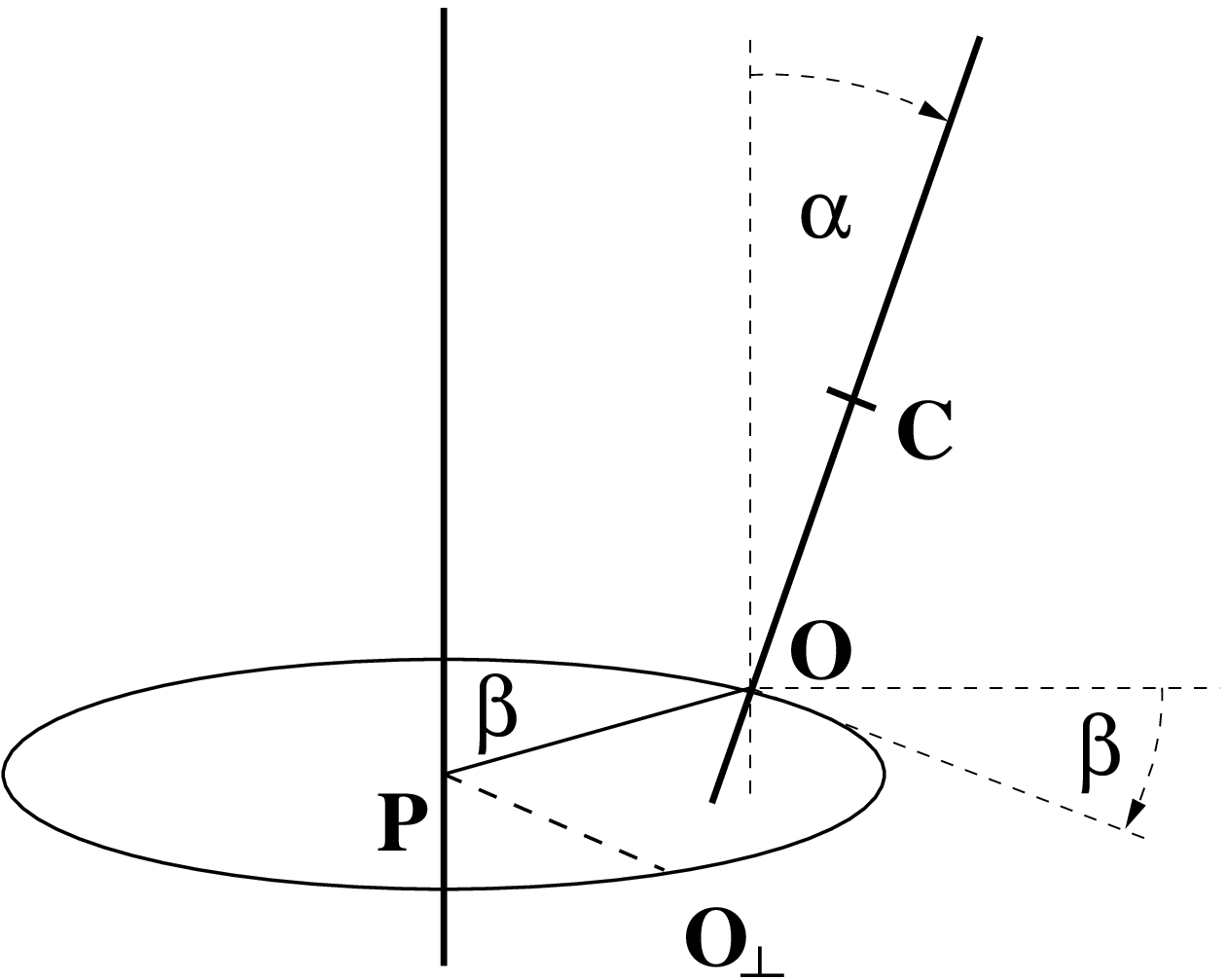}
\caption{(left) Sketch of a rod with center of mass (C) and
  orientation $\alpha$ intersecting a arbitrary unit line (dashed) to
  illustrate the calculation of the number of intersecting rods
  $\rho_{\rm MC}$. (right) Two rods with minimal distance $R$ are
  considered to be mutually interacting only if their line of closest
  approach ($\overline{OP}$) is orthogonal to both. $O_\perp$ only
  serves to illustrate the analogy to the two-dimensional setup (see
  text).}
\label{fig:statistics}
\end{figure}

We derive the
radial density of obstacles that effectively hinder the test rods
fluctuations. To this end we consider the test rod to be aligned along
the z-axis without loss of generality (see Fig.~\ref{fig:statistics}
(right)). As criterion for effective obstruction of transverse
fluctuations between test rod and obstacle, we demand that the line
connecting their points of closest approach ($\overline{PO} $) is
orthogonal to both polymers. As a projection of the obstacle to the
plane spanned by the test rod and $\overline{PO}_\perp$ (dashed)
recovers the setup discussed in Appendix \ref{sec:density}, the
coordinates of the center of mass (C) can readily be extended to three
dimensions by the radial distance $R$ and the angle $\beta$ to
$r=(\sin(\alpha)s \sin(\beta)-\cos(\beta)R, \sin(\alpha)s \cos(\beta)
+ \sin(\alpha) R, x-\cos(\alpha)s)$ and with the Jacobian determinant
$|\partial r/\partial(x,s,R)|=\sin(\alpha)$ the integration gives
\begin{eqnarray}  \label{eq:app1_2}
\rho(R)&=& \frac{2 \nu}{4 \pi}\int_{-\frac{L}{2}}^{\frac{L}{2}} dx \int_{-L/2}^{L/2} ds
\int_{0}^{\pi} d\alpha \int_0^{2 \pi} d\beta
\sin^2(\alpha) \nonumber\\
&=& \frac{\pi}{2} \nu L^2 \;.
\end{eqnarray}
Consequently, the density seen by an stiff segment of length $\bar L$ in
the IRM will be 
\begin{equation}
\rho=\frac{\bar L}{L} \frac{\pi}{2}\nu L^2= \frac{\pi}{2} \nu L \bar L
\;.
\end{equation}
Note that in the simplified geometry the obstacle density is not a
complete radial density but a line density of obstacles on one of the
four axes (positive and negative $y$ and $z$ axis). To recover the
complete radial density, one has to sum over all of these. Hence the
obstacle density on either one of the four axes has to be:
\begin{equation}
\rho = \frac{\pi}{2} \frac{\nu L \bar L}{4} \;.
\end{equation}

\section{Mode analysis of polymer and tube} \label{sec:mode-analys-polym}

Using the dimensionless arc-length $\tilde s=s/L$, polymer conformation
$\tilde y(\tilde s)=y(\tilde s L)/L$, tube center $\tilde y^0(\tilde
s)=y^0(\tilde s L)/L$ and persistence length $\eta=l_{\rm p}/L$ the
Hamiltonian (\ref{eq:hamilton}) can be written for one transverse
coordinate as
\begin{equation} \label{eq:dimless_hamilton}
\beta H = \frac{\eta}{2} \int_0^1 d \tilde s \left[ (\partial_{\tilde s}^2
  \tilde y)^2 + l_{\rm d}^{-4}(\tilde y - \tilde y^0)^2 \right] \;.
\end{equation}
For free boundary conditions the Hamiltonian is diagonalized by an
orthonormal set of eigenfunctions $\psi_k \approx sin(q_k \sigma +
\phi_k)$ with $q_k \approx \pi (k+1/2)$ \cite{aragon85}. Expanding
both polymer and tube center in modes as $\tilde y(\sigma)=\sum_k y_k
\psi_k$ and $\tilde y^0(\sigma)=\sum_k y_k^0 \psi_k$ and using the
orthogonality of the eigenfunctions allows one to write the
Hamiltonian in the suggestive form
\begin{eqnarray}
\beta H &=& \frac{\eta}{2} \sum_k (q_k^4+l_{\rm d}^{-4})\left(y_k-\frac{y_k^0}{1+(q_k
      l_{\rm d})^4}\right)^2 \nonumber\\
 && +\frac{q_k^4 {y_k^0}^2}{1+\left(q_k l_{\rm d} \right)^4}  \;.
\end{eqnarray}
By comparison to (\ref{eq:dimless_hamilton}) one can read of the
minimum of the confinement potential, i.e. the average tube center:
\begin{equation} \label{eq:app2_1}
  \overline{y_k}=y_k^0/(1+q_k^4l_{\rm d}^4) \;.
\end{equation}

Additionally, this allows to write the complete transverse
fluctuations as a sum over the inverse confinement strength of all modes: 
\begin{equation} \label{eq:app2_2}
L_\perp^2 := \frac{1}{L} \int_0^L ds
\overline{(x(s)-\overline{x(s)})^2}=\frac{L^2}{\eta} \sum_k
\frac{1}{q_k^4+l_{\rm d}^{-4}} \;.
\end{equation}
For the dimensionless function $h(l_{\rm d})$ relating the tube diameter to
the deflection length follows:
\begin{equation} \label{eq:app2_3}
h(ld)=\sum_k \frac{1}{q_k^4+l_{\rm d}^{-4}} \;.
\end{equation}

\bibliographystyle{unsrt}
\bibliography{version1}

\end{document}